\documentclass[man]{apa7}
\usepackage{amsmath}
\usepackage{graphicx}
\usepackage{lipsum}
\usepackage[final]{changes}
\usepackage[style=numeric,sorting=nyt]{biblatex}
\newcommand{\hl}[1]{#1}
\DeclareLanguageMapping{english}{english-apa}

\title{On the Design of Corrugated Boards: A New FEM Modeling and Experimental Validation}
\shorttitle{A New FEM Modeling and Experimental Validation}
\author{Ricardo Fitas\textsuperscript{1}, Heinz Joachim Schaffrath\textsuperscript{1}, Samuel Schabel\textsuperscript{1}}

\authorsaffiliations{
\textsuperscript{1}Technical University of Darmstadt \\
Contact: ricardo.fitas@tu-darmstadt.de, ORCID: 0000-0001-5137-2451 \\
heinz-joachim.schaffrath@tu-darmstadt.de, ORCID: 0000-0002-3250-6618 \\
samuel.schabel@tu-darmstadt.de, ORCID: 0000-0002-7207-8208
}

\abstract{
This study presents a simplified FEM modeling approach suitable for large structures made of corrugated boards, such as customized packages, based on a homogenization method, which is combined with correction factors for internal mechanisms. The homogenization process reduces computational time by transforming flute geometries into equivalent elastic models. In large deformations and in the presence of contact for a given geometry, the effective elastic modulus in the thickness direction, as well as the effective thickness of the structure, are corrected by two statistical Weibull distributions representing the contact and buckling mechanisms in a corrugated board. The Weibull parameters are obtained via experimental analysis, and such a process is then validated. The results demonstrate that the statistical parameters ($\beta_1 = 0.14$, $\beta_2 = 1.31$) can be used for the simplistic representation of corrugated boards, being computationally efficient. This research contributes to the optimization of corrugated packaging design, specifically by simplifying FEM models for faster yet equally accurate simulations.
}


\begin{document}

\maketitle

\section{Introduction}

The corrugated board market has witnessed significant advancements driven by both technological innovations and a growing emphasis on sustainability. Corrugated boards, known for their unique structure with wavy flute layers, offer a combination of strength, cushioning, and environmental benefits, making them a preferred choice in the packaging industry \parencite{grabski2023identification}. The mechanical properties of corrugated boards are influenced by various factors, including the type of paper used, the geometry of the flute, and environmental conditions such as humidity and temperature, which can affect their stiffness and strength \parencite{cornaggia2023influence}. Recent studies have focused on optimizing the design and production of corrugated boards through advanced numerical modeling. Numerical modeling, particularly finite element analysis, has been employed to simulate the structural performance of corrugated boards, allowing for precise predictions of their mechanical behavior under different loading conditions \parencite{aduke2023analysis}. 

Finite Element Method (FEM) models for corrugated board structures, such as boxes, face significant challenges due to the intricate geometry of the flute and the material non-linearity inherent in the paperboard. The complexity of the flute geometry necessitates a high computational cost, which is often impractical when the primary interest lies in the overall structural geometry rather than the detailed internal structure. Park et al. explored the bending behavior of corrugated boards using FEM, noting that while the cross-machine direction (CD) bending stiffness was well-simulated, the machine-direction (MD) behavior posed challenges due to convergence issues and contact condition variability \parencite{park2024finite}.

Further research by Cillie and Coetzee identified discrepancies in FEM predictions of in-plane deformation, suggesting that incorporating initial imperfections and adopting non-linear material models could enhance model accuracy \parencite{cillie2022experimental}. The use of image analysis and genetic algorithms to identify geometric features of corrugated boards, as presented by Grabski and Garbowski, offers a promising avenue for improving the precision of FEM models by providing accurate geometric inputs \parencite{grabski2023identification}. Additionally, the work by Di Russo et al. on evaluating different wave configurations in corrugated boards underscores the importance of considering various flute geometries and their impact on mechanical performance, which can inform the development of more efficient FEM models \parencite{russo2023evaluation}.

Current research highlights the need for simplified modeling approaches that can accurately predict the mechanical behavior of corrugated boards while reducing computational demands \parencite{fitas2023review}. 

Homogenization techniques have been proposed as a viable solution to the computational challenges posed by the complex geometry of corrugated boards. For instance, Aduke et al. developed a homogenized FEM model that effectively predicted the failure load of corrugated boxes under compression, achieving a minimal variation of 0.1\% compared to experimental results, thus demonstrating the potential of simplified models in capturing essential structural responses without excessive computational resources \parencite{aduke2023analysis}.  \hl{Moreover, Marek and Garbowski} \parencite{marek2017homogenization} \hl{compared two major methodologies — the classical laminated plate theory (CLPT) and the deformation energy equivalence method (DEEM) — for corrugated sandwich structures and demonstrated that both approaches can yield consistent estimates of effective stiffness in the elastic regime. Similarly, Starke} \parencite{starke2020corrugated} \hl{developed a finite element homogenization framework for corrugated carton panels, modeling the core as an equivalent homogeneous layer. These efforts provide the theoretical basis for the current compliance-based homogenization strategy, which is extended in our work by including orientation-aware transformations and probabilistic failure characterization.}

Greco also explores the integration of contact mechanics and proposes a novel micro-mechanical approach that incorporates fracture mechanics concepts to study micro-cracking in composites. This approach uses a J-integral formulation to develop non-linear macroscopic constitutive laws that account for changes in micro-structural configuration due to crack growth and contact, \parencite{greco2009homogenized}.

Additionally, Wriggers and Nettingsmeier emphasize the importance of finite element analysis in investigating contact behavior at the micro-mechanical level, which can replace classical statistical methods and lead to homogenized constitutive models for contact. Their work demonstrates the potential of computational tools to derive macroscopic constitutive equations that incorporate the behavior of different scales, particularly in the context of rough surface interactions \parencite{wriggers2007homogenization}.

\hl{Moreover, previous models for corrugated sandwich panels have extensively investigated the contribution of out-of-plane shear stiffness to overall mechanical behavior. In particular, Nordstrand et al.} \parencite{nordstrand1994transverse} \hl{and Carlsson et al.} \parencite{carlsson2001elastic} \hl{demonstrated that the transverse shear modulus is highly sensitive to the core geometry and damage state, and highlighted its dominant role in panel response under bending and more complex loading scenarios. Their works employed both analytical and finite element techniques, often within the framework of first-order shear deformation theory, to capture stiffness degradation due to delamination and core shape distortion.}

Despite these advancements, the challenge of scale separability persists, as highlighted by Kerfriden et al., who discuss the limitations of homogenization methods when a clear separation of scales is absent. They propose a method to quantify scale separability in stochastic homogenization, which is crucial for accurately predicting the overall behavior of structures under varying macroscopic gradients \parencite{kerfriden2014certification}. 
The integration of homogenization processes in FEM, considering also simplified parameters accounting for contact mechanics and buckling, is missing.

In this paper, homogenization is integrated with common non-linear phenomena in corrugated boards, such as buckling and contact between flute and liners, to simplify FEM models while maintaining non-linear characteristics for faster simulations. This approach is crucial not only to achieve faster simulations but also for reliable analysis in optimization routines, as it balances the need for detailed mechanical behavior representation with computational efficiency. 

The remainder of the paper is structured as follows: the next section details the materials and methods used in the study, including the specific numerical homogenization techniques applied, the modeling extension to other deformed flute geometries, the determination of elastic moduli involves the following standardized tests: the Edgewise Crush Test (ECT) \parencite{iso3037}, the Short-Span Compression Test (SCT) \parencite{iso9895}, and the Flat Crush Test (FCT) \parencite{iso3035}. These parameters are considered crucial for modeling the mechanical behavior of corrugated boards. The section "Results and Discussion" presents the outcomes of all the simulations and compares them with experimental data to evaluate the effectiveness of the homogenization approach. The discussion highlights the advantages of the proposed methodology in terms of computational efficiency and accuracy, as well as its limitations and potential areas for improvement. Finally, the conclusions summarize the key findings of the study, emphasizing the practical implications of the research for the design and optimization of corrugated board packaging and suggesting directions for future research.

\section{Materials and Methods}

\subsection{Homogenization Method}

\begin{figure}[h]
    \centering
    \includegraphics[width=\textwidth]{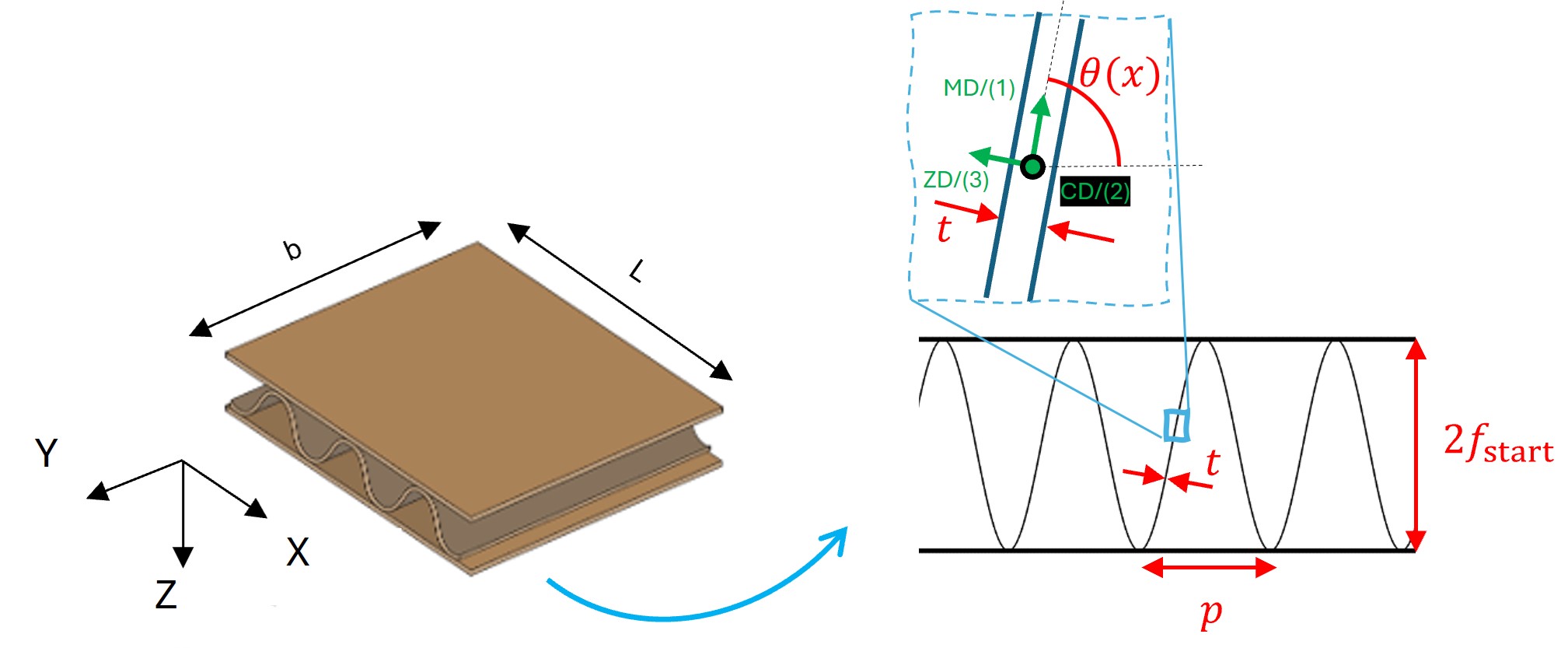}
    \caption{Definition of the different stiffness directions (local and global coordinates).}
    \label{fig01}
\end{figure}

The homogenization process, as described in \parencite{aduke2023analysis, fitas2025corrugated,starke2020corrugated}, is used to compute the effective elastic modulus ($E_{z, \text{eff}}$) and thickness ($th_{\text{eff}}$) of corrugated boards by considering the geometric and material properties of the flute structure. In this study, the process follows several key steps, including defining the material compliance matrix and performing transformations to account for the flute's geometry. The material properties are named according to the naming of the axes in Figure \ref{fig01}.

\subsubsection{Material Compliance Matrix}

The compliance matrix for the corrugated board is defined based on the elastic properties of the material, including Young's moduli ($E_1$, $E_2$, and $E_3$), the shear moduli ($G_{12}$, $G_{13}$, and $G_{23}$), and the Poisson ratios ($\nu_{12}$, $\nu_{13}$, and $\nu_{23}$). The general form of the compliance matrix ($\mathbf{C}^{(123)}$) in the material coordinate system is given by (\ref{eq:C123}).

\begin{equation} \label{eq:C123}
\mathbf{C}^{(123)} = \begin{bmatrix}
    \frac{1}{E_1} & -\frac{\nu_{12}}{E_1} & -\frac{\nu_{13}}{E_1} & 0 & 0 & 0 \\
    -\frac{\nu_{12}}{E_1} & \frac{1}{E_2} & -\frac{\nu_{23}}{E_2} & 0 & 0 & 0 \\
    -\frac{\nu_{13}}{E_1} & -\frac{\nu_{23}}{E_2} & \frac{1}{E_3} & 0 & 0 & 0 \\
    0 & 0 & 0 & \frac{1}{G_{12}} & 0 & 0 \\
    0 & 0 & 0 & 0 & \frac{1}{G_{13}} & 0 \\
    0 & 0 & 0 & 0 & 0 & \frac{1}{G_{23}}
\end{bmatrix}
\end{equation}

\subsubsection{Flute Geometry and Transformation}

The corrugated structure is modeled as a periodic sinusoidal profile, described by (\ref{eq:H}).

\begin{equation} \label{eq:H}
H(x) = \frac{f_{\text{start}}}{2} \sin \left( \frac{2 \pi}{p} x \right)
\end{equation}

where $f_{\text{start}}$ is the initial flute height, and $p$ is the wavelength of the flute. The profile's rotation angle $\theta$ at any point along the x-axis is given by (\ref{eq:theta}).

\begin{equation} \label{eq:theta}
\theta(x) = \tan^{-1} \left( \frac{\pi f_{\text{start}}}{p} \cos \left( \frac{2 \pi}{p} x \right) \right)
\end{equation}

To account for the periodic nature of the flutes, transformation matrices are applied to the compliance matrix. The transformation matrices for rotation about the y-axis are defined in (\ref{eq:T_e}).

\begin{equation} \label{eq:T_e}
\mathbf{T} = \begin{bmatrix}
    c^2 & 0 & s^2 & 0 & -sc & 0 \\
    0 & 1 & 0 & 0 & 0 & 0 \\
    s^2 & 0 & c^2 & 0 & sc & 0 \\
    0 & 0 & 0 & c & 0 & -s \\
    2sc & 0 & -2sc & 0 & c^2 - s^2 & 0 \\
    0 & 0 & 0 & -s & 0 & c
\end{bmatrix}
\end{equation}

where $c = \cos(\theta)$ and $s = \sin(\theta)$. The transformed compliance matrix in the global coordinate system ($\mathbf{C}^{(xyz)}$) is calculated as in (\ref{eq:Cxyz}).

\begin{equation} \label{eq:Cxyz}
\mathbf{C}^{(xyz)} = \mathbf{T} \mathbf{C}^{(123)} \mathbf{T}^{T}
\end{equation}

where $\mathbf{T}^{T}$ is the transpose of $\mathbf{T}$.

\subsubsection{Homogenization Along the Thickness}

The homogenization method involves integrating the material properties through the thickness of the fluting structure. The vertical thickness of the fluting, $t_v$, is calculated as in (\ref{eq:tv}).

\begin{equation} \label{eq:tv}
t_v = \frac{t}{\cos(\theta)}
\end{equation}

where $t$ is the thickness of the liner, the effective elastic modulus, and effective thickness are computed by integrating the stiffness matrices $\mathbf{A}$ (\ref{eq:A}) and $\mathbf{D}$ (\ref{eq:D}) over the flute's periodic length $p$.

\begin{equation} \label{eq:A}
\mathbf{A} = \frac{1}{p} \int_0^p t_v \mathbf{Q} \, dx
\end{equation}

\begin{equation} \label{eq:D}
\mathbf{D} = \frac{1}{p} \int_0^p (H^2(x) t_v + \frac{t_v^3}{12}) \mathbf{Q} \, dx
\end{equation}

where $\mathbf{Q}$ (\ref{eq:Q}) is a reduced stiffness component derived from the transformed compliance matrix.

\begin{equation} \label{eq:Q}
\mathbf{Q} = \begin{bmatrix}
    \mathbf{C}^{(xyz)}_{11} & \mathbf{C}^{(xyz)}_{12} & \mathbf{C}^{(xyz)}_{14} \\
   \mathbf{C}^{(xyz)}_{21} & \mathbf{C}^{(xyz)}_{22} & \mathbf{C}^{(xyz)}_{24} \\
    \mathbf{C}^{(xyz)}_{41} & \mathbf{C}^{(xyz)}_{42} & \mathbf{C}^{(xyz)}_{44}
\end{bmatrix}^{-1}
\end{equation}

\subsubsection{Effective Thickness and Elastic Modulus}

The final step in the homogenization process is calculating the effective thickness ($th_{\text{eff}}$) and the effective elastic modulus ($E_{z, \text{eff}}$). The effective thickness is calculated using (\ref{eq:theff}).

\begin{equation} \label{eq:theff}
th_{\text{eff}} = \sqrt{\frac{12 \left( \mathbf{D}_{11} + \mathbf{D}_{22} + \mathbf{D}_{33} \right)}{\mathbf{A}_{11} + \mathbf{A}_{22} + \mathbf{A}_{33}}}
\end{equation}

The effective elastic modulus is then determined by (\ref{eq:Ezeff}) \parencite{aduke2023analysis,marek2017homogenization}.

\begin{equation} \label{eq:Ezeff}
E_{z, \text{eff}} = \frac{12}{{th_{eff}}^3 \mathbf{D^{-1}}_{33}}
\end{equation}

This homogenization process allows the prediction of the effective mechanical properties of the corrugated board, considering the geometry and material characteristics, enabling the optimization of the board's design.

\subsubsection{Modeling Extension to different Out-of-Plane Compression Phases}

\hl{When the corrugated board is compressed in the out-of-plane direction, its flute undergoes significant deformation.} The initial geometry described by $H(x)$ can no longer be represented by a single continuous function. Instead, the deformation introduces multiple overlapping geometries, each characterized by its own periodicity. This results in the need to redefine the geometry and stiffness matrices for the deformed state.

\begin{figure}[h]
    \centering
    \includegraphics[width=\textwidth]{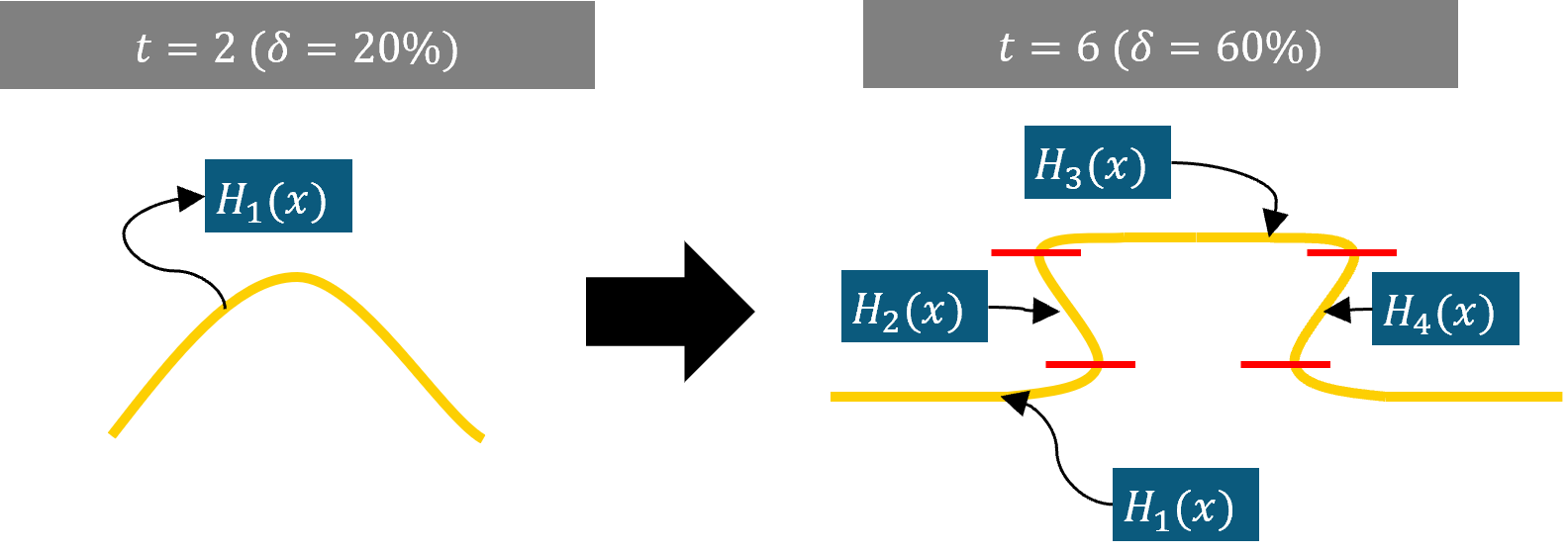}
    \caption{Left: Definition of a single function before buckling vs. Right: Definition of different functions after buckling.}
    \label{fig02}
\end{figure}

The new deformed geometry, at the deformation time step $t$, is determined by calculating the arc length of the deformed curve while preserving material continuity. The geometric parameters and the deformation input are used to solve an optimization problem, ensuring that the arc-length constraint is satisfied while finding the new geometry. This results in a set of multiple periodic functions $H_i(x)$ for the different deformed regions (Figure \ref{fig02}).

The modified direction functions $\theta_i(x, t)$ for the overlapping geometries are generally calculated in (\ref{eq:thetaixt}).

\begin{equation} \label{eq:thetaixt}
\theta_i(x,t) = \tan^{-1}\left( \frac{\partial H_i(x,t)}{\partial x}  \right)
\end{equation}

The stiffness matrices $A$ and $D$ must also be adjusted to account for the multiple overlapping regions. The updated expression for $\mathbf{A}$ (\ref{eq:A11}) and $\mathbf{D}$ (\ref{eq:D11}) becomes a summation over all deformed regions (Note: this expression can also be adapted and applied for the remaining components of the stiffness matrices).

\begin{equation} \label{eq:A11}
\mathbf{A}^{(t)} = \sum_{i=1}^{n_t} \mathbf{A}^{(i,t)},
\end{equation}

\begin{equation} \label{eq:D11}
\mathbf{D}^{(t)} = \sum_{i=1}^{n_t} \mathbf{D}^{(i,t)},
\end{equation}

where $n$ is the number of overlapping regions.

\hl{It is important to note that the current homogenization model is not intended to explicitly capture local buckling shapes or individual instability modes of the flute structure. Instead, the goal is to approximate the average out-of-plane response of the board over a representative length, incorporating its geometric and material anisotropy.}

\subsection{Methodology Overview}
\begin{figure}[h]
    \centering
    \includegraphics[width=\textwidth]{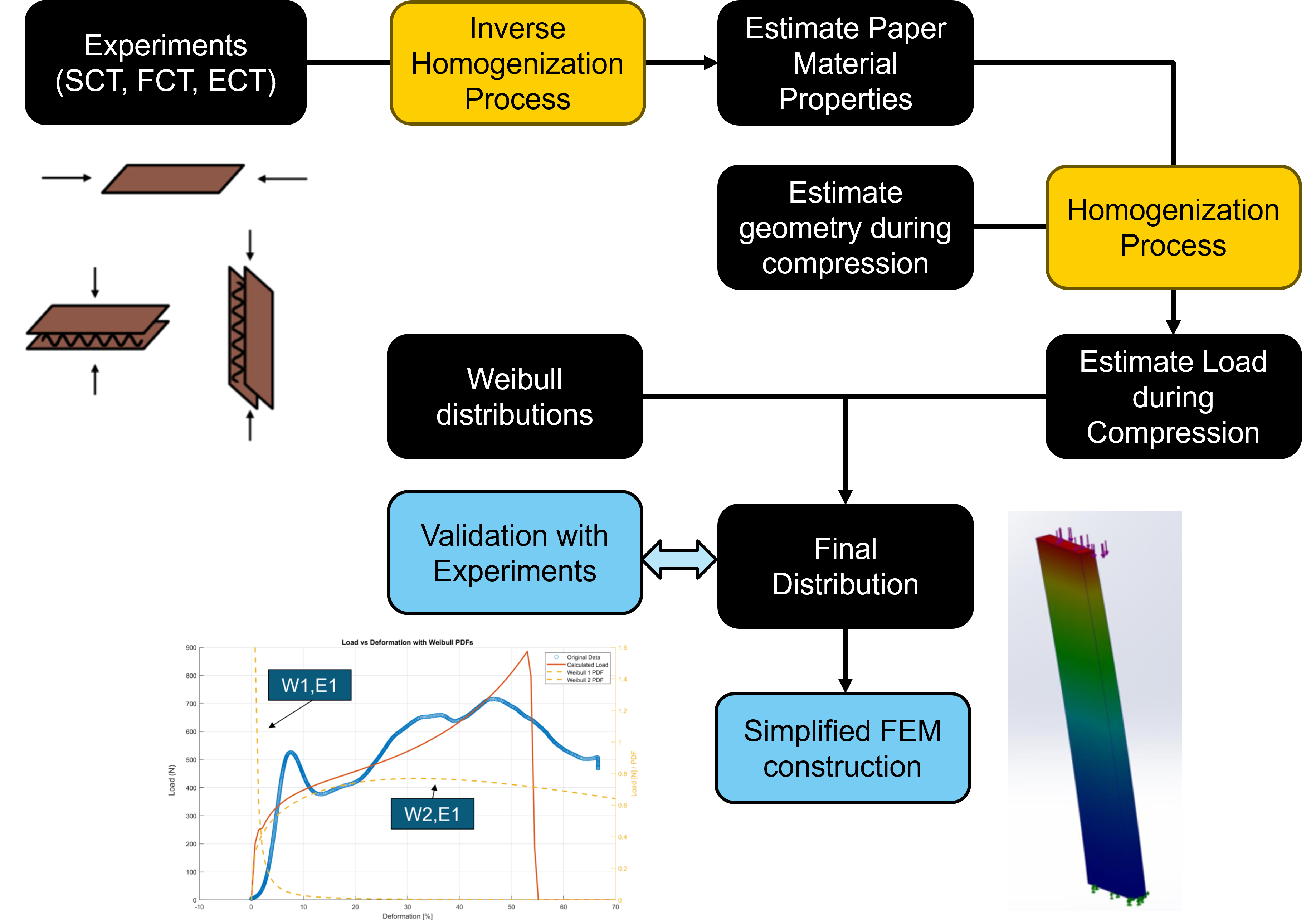}
    \caption{Theoretical Framework being applied for this study.}
    \label{fig1}
\end{figure}

The methodology followed a structured approach to characterize and model the mechanical behavior of corrugated boards, as shown in Figure \ref{fig1}. It starts with the experimental tests, including ECT, SCT, and FCT, to obtain baseline mechanical properties for the paper material. \hl{The experimental tests are conducted directly on the corrugated boards. The resulting experimental values are then used in an inverse homogenization process to estimate effective, homogenized material properties (namely out-of-plane elastic modulus and effective thickness) from measured global behavior. This is achieved by adjusting the material parameters until the numerical output matches the experimental results, thus deducing the equivalent properties of the elastic properties of the paper.} After estimating the material properties of the paper, $E_{MD}$, $E_{CD}$, $E_{z}$, the homogenization \hl{process, described in the previous subsection, for different out-of-plane compression phases is conducted. This process results in an estimated load-deformation curve. In addition to this,} two Weibull distributions that describe failure mechanisms (contact and buckling) are combined with the necessary compressive load, and their parameters are estimated. Finally, statistical tests validated the reliability of the model against a different sample. The Weibull parameters, as well as the simplified model, resulted in the simplified FEM construction. 

\subsection{Estimation of the Young's Moduli for the Paper Material}

\begin{figure}[h]
    \centering
    \includegraphics[width=\textwidth]{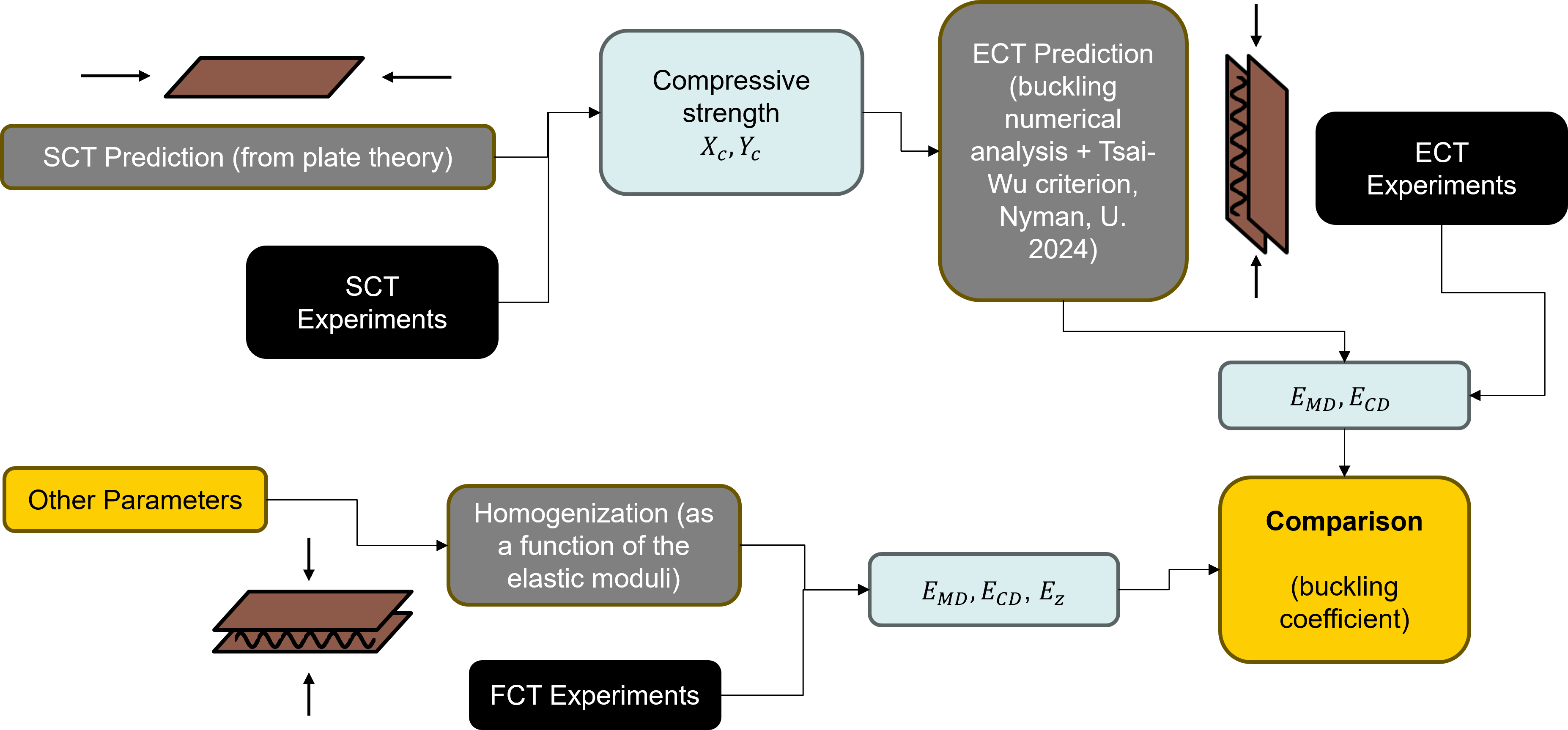}
    \caption{Framework being used for the estimation of the Young's Moduli for the paper material.}
    \label{fig2}
\end{figure}

\hl{The present section addresses the estimation of the intrinsic material moduli of the paper layers themselves. These include the in-plane moduli $E_{MD}$ and $E_{CD}$, and the through-thickness modulus $E_{ZD}$, which are required as inputs to the compliance matrix used in the homogenization. This is achieved using an inverse estimation approach based on SCT and ECT experiments, as outlined in Figure} \ref{fig2}.

\hl{The moduli $E_{MD}$ and $E_{CD}$ are estimated from in-plane experimental data using the SCT and ECT results.} From the SCT experiments, the compressive strength can be measured. This compressive strength serves as the input for the estimation of the moduli using combined ECT experiments from industrial tests \parencite{schmitt2019fibre}. \hl{The methodology employed follows the critical stress analysis conducted by} \parencite{nyman2004continuum}. \hl{As shown in Figure} \ref{fig2} \hl{, the ECT prediction incorporates a buckling-based numerical analysis combined with the Tsai-Wu failure criterion, which allows capturing multi-axial stress interactions in the liner and flute layers. For further details on the implementation of the Tsai-Wu criterion and the in-plane critical stress analysis used in this step, the reader is referred to} \parencite{nyman2004continuum}.

\hl{A numerical calibration process is employed to identify the set of elastic moduli that best reproduce the experimental ECT values. This is done by minimizing the absolute error between the simulated and observed results.}

\hl{Separately, the out-of-plane modulus $E_{ZD}$ is estimated using a homogenization-based inverse modeling approach that incorporates experimental FCT data. The homogenized effective stiffness $E_{z,\text{eff}}$ is treated as a function of $E_{ZD}$ and the other material parameters, which are adjusted until the model reproduces the observed out-of-plane behavior.}

\hl{The optimized values of $E_{MD}$, $E_{CD}$, and $E_{ZD}$ are then inserted into the compliance matrix $\mathbf{C}^{(123)}$ used in the homogenization model, allowing for the computation of $E_{z,\text{eff}}$ and $th_{\text{eff}}$.}

Finally, a comparison between the results is conducted. Since a significant difference between values is anticipated, it is assumed that this discrepancy arises from the absence of the buckling coefficient necessary for accurate estimation in the in-plane analysis.

\hl{To quantify this, a buckling correction factor is introduced based on the observed discrepancy. This correction factor accounts for geometric imperfections, local buckling initiation, and nonlinear deformation effects that are not captured by classical continuum models. It ensures that the estimated in-plane stiffness better reflects the actual behavior observed in compression tests. Neglecting this correction would lead to a poor fit between modeled and experimental data, and may cause systematic error in the compliance matrix and subsequent homogenization calculations.}

\subsection{Weibull Distribution in Compression Analysis of Corrugated Boards}

The \textbf{Weibull distribution} is widely used for reliability analysis and failure modeling, describing the probability of failure of a material over time. It is characterized by two main parameters:

\begin{itemize}
    \item \textbf{Shape parameter ($\beta$)}: Defines the failure rate behavior:
    \begin{itemize}
        \item If $\beta < 1$: Failures are caused by early-stage defects or manufacturing issues.
        \item If $\beta \approx 1$: Failures occur randomly over time.
        \item If $\beta > 1$: Failures occur progressively.
    \end{itemize}
    \item \textbf{Scale parameter ($\eta$)}: Represents the characteristic life, where 63.2\% of failures occur by time $\eta$.
\end{itemize}

The probability density function (PDF) of the Weibull distribution is given by (\ref{eq:ft}).

\begin{equation} \label{eq:ft}
f(t) = \frac{\beta}{\eta}\left(\frac{t}{\eta}\right)^{\beta-1} e^{-(t/\eta)^\beta}
\end{equation}
where $f(t)$ is the failure rate at time $t$.

\subsubsection{Modeling Approach}
\begin{figure}[h]
    \centering
    \includegraphics[width=0.7\textwidth]{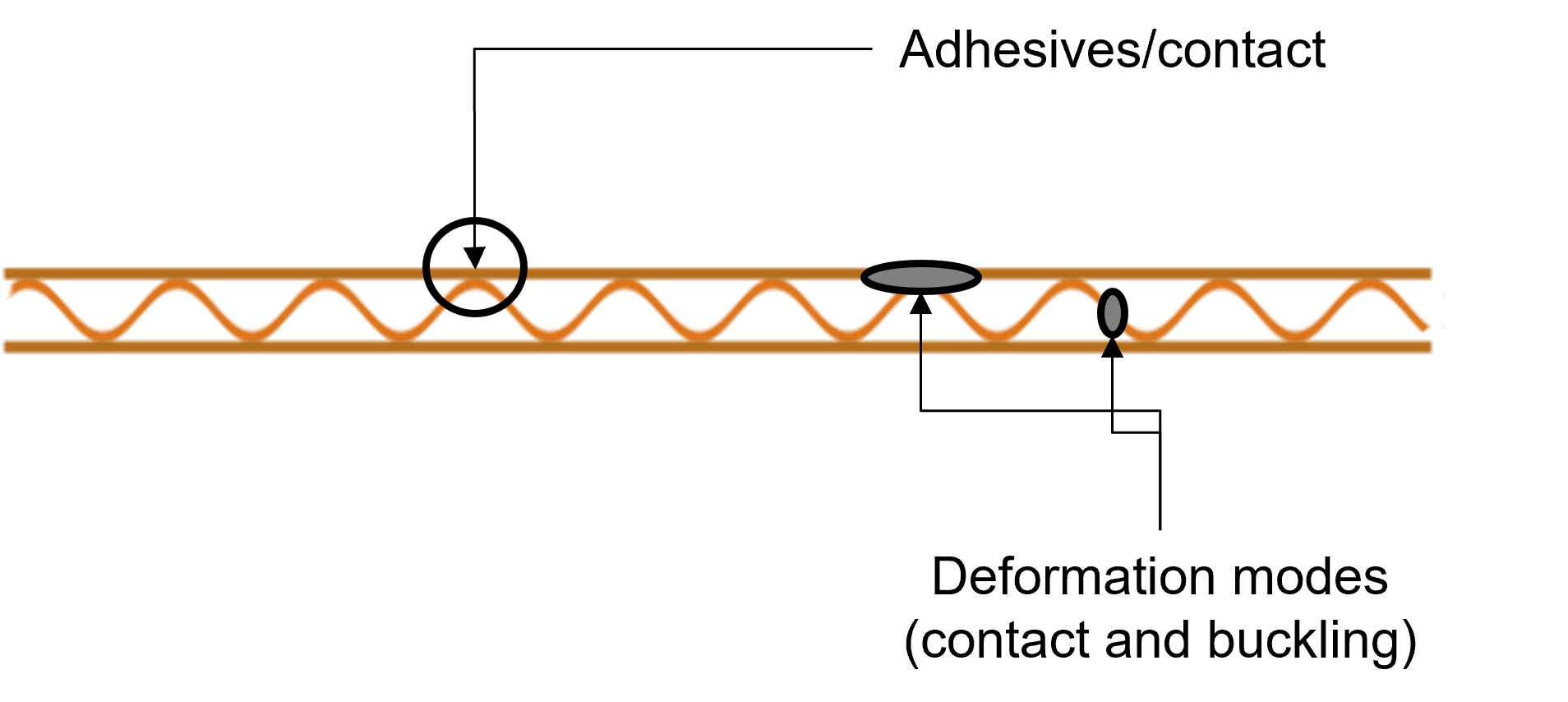}
    \caption{Schematic representation of the dominant deformation modes and associated damage mechanisms during compressive loading.}
    \label{fig3}
\end{figure}

The failure modes in the \textbf{horizontal} and \textbf{vertical} directions can be modeled using two independent Weibull distributions. Each direction captures a distinct deformation mechanism, as illustrated in Figure \ref{fig3}. \hl{It illustrates the two directions of deformation and their associated damage mechanisms: vertical flute buckling under out-of-plane compressive loads and localized settling or softening in the horizontal direction. The latter is not treated as a classical structural failure, but rather as an energy-dissipating mechanism attributed to contact loss at the flute-liner interfaces. Therefore,} the horizontal direction is assumed to capture primarily the contact between layers, and, due to that phenomenon, deformation happens with less load. The vertical direction is assumed to capture primarily the vertical buckling of the flute.
\hl{While the contact loss mechanism is modeled as an in-plane Weibull process for simplicity, it may arise from mixed-mode loading, especially in the presence of imperfections. Nevertheless, its initiation is typically associated with early-stage in-plane compressive settling.}

\hl{To still account for local failure phenomena — particularly flute buckling and contact degradation — the two statistical Weibull distributions are superimposed onto the homogenized response. These distributions are calibrated against experimental compression data and provide a means to incorporate localized nonlinear effects without abandoning the computational efficiency of the homogenized model.}

\hl{The probability density function $f(t)$ is not used to model failure at the material point level, but rather as a statistical weighting applied to the global load–deformation response predicted by the homogenized elastic model. Specifically, two Weibull distributions are introduced: one corresponding to early contact degradation between flutes and liners, and another representing stochastic buckling events in the vertical direction. These are calibrated against experimental compression data.}

\hl{The function $f(t)$, governed by the shape parameter $\beta$ and scale parameter $\eta$, is used to compute a cumulative failure probability that modulates the simulated load curve. Low values of $\beta$ (e.g., $\beta < 1$) imply gradual onset of failure, while higher values ($\beta > 1$) indicate sharper, more stochastic failure transitions. The resulting load response is given by the product of the elastic prediction and a softening factor derived from the Weibull function. This approach captures energy dissipation and stiffness degradation observed in experiments without explicitly modeling geometric or material nonlinearity.}

\section{Results and Discussion}

The material properties of the paper material were determined using ECT and SCT experiments. The results for Young's moduli in the machine direction (MD) and cross direction (CD) are summarized in Table~\ref{tab:ect_results}. \hl{It summarizes the measured mechanical properties for three standard flute types commonly used in corrugated board manufacturing: B, C, and E. These flute classifications correspond to different geometrical profiles as defined in DIN 55468-1} \parencite{DIN55468-1_2021}. \hl{The B-flute has a medium thickness and high flute density, C-flute is thicker and coarser, and E-flute is thinner and more compact, each offering different trade-offs between stiffness and cushioning. Different types of paper materials are also classified as recycled, non-recycled, or mixed.}
One can observe that non-recycled flutes are much stronger in MD compared to other materials. Moreover, the mixed type-E flute presents the weakest properties in such a direction, but recycled materials can also present lower stiffness values. In CD, recycled C-Flute presents the highest stiffness, while non-recycled flutes can be weaker compared to other types.

\begin{table}[H]
    \centering
    \caption{Estimated Young's Moduli in MD and CD for different flute types.}
    \label{tab:ect_results}
    \begin{tabular}{lcccc}
        \toprule
        \textbf{Material / Flute Type} & \textbf{Mean} $E_{MD}$ [GPa] & \textbf{SD} $E_{MD}$ [GPa] & \textbf{Mean} $E_{CD}$ [GPa] & \textbf{SD} $E_{CD}$ [GPa] \\
        \midrule
        Mixed* B-Flute & 1.68 & 0.54 & 1.10 & 0.37 \\
        Mixed* C-Flute & 1.69 & 0.76 & 1.15 & 0.57 \\
        Mixed* E-Flute & \textbf{1.46***} & 0.35 & 1.14 & 0.23 \\
        Recycled B-Flute & 1.59 & 0.85 & 1.12 & 0.36 \\
        Recycled C-Flute & 1.47 & 0.53 & \textbf{1.20**} & 0.60 \\
        Recycled E-Flute & 1.77 & 0.58 & 1.13 & 0.62 \\
        Non-Recycled B-Flute & 1.75 & 0.44 & \textbf{1.08***} & 0.21 \\
        Non-Recycled C-Flute & \textbf{1.89**} & 0.90 & 1.11 & 0.51 \\
        \bottomrule
    \end{tabular}

    *Mixed means part of the paper is recycled and another part is non-recycled.
    **Highest values by columns (p<0.01, Kruskal-Wallis, Welch's ANOVA)
    ***Lowest values by columns (p<0.01, Kruskal-Wallis, Welch's ANOVA)
\end{table}

\begin{figure}[H]
    \centering
    \includegraphics[width=0.7\textwidth]{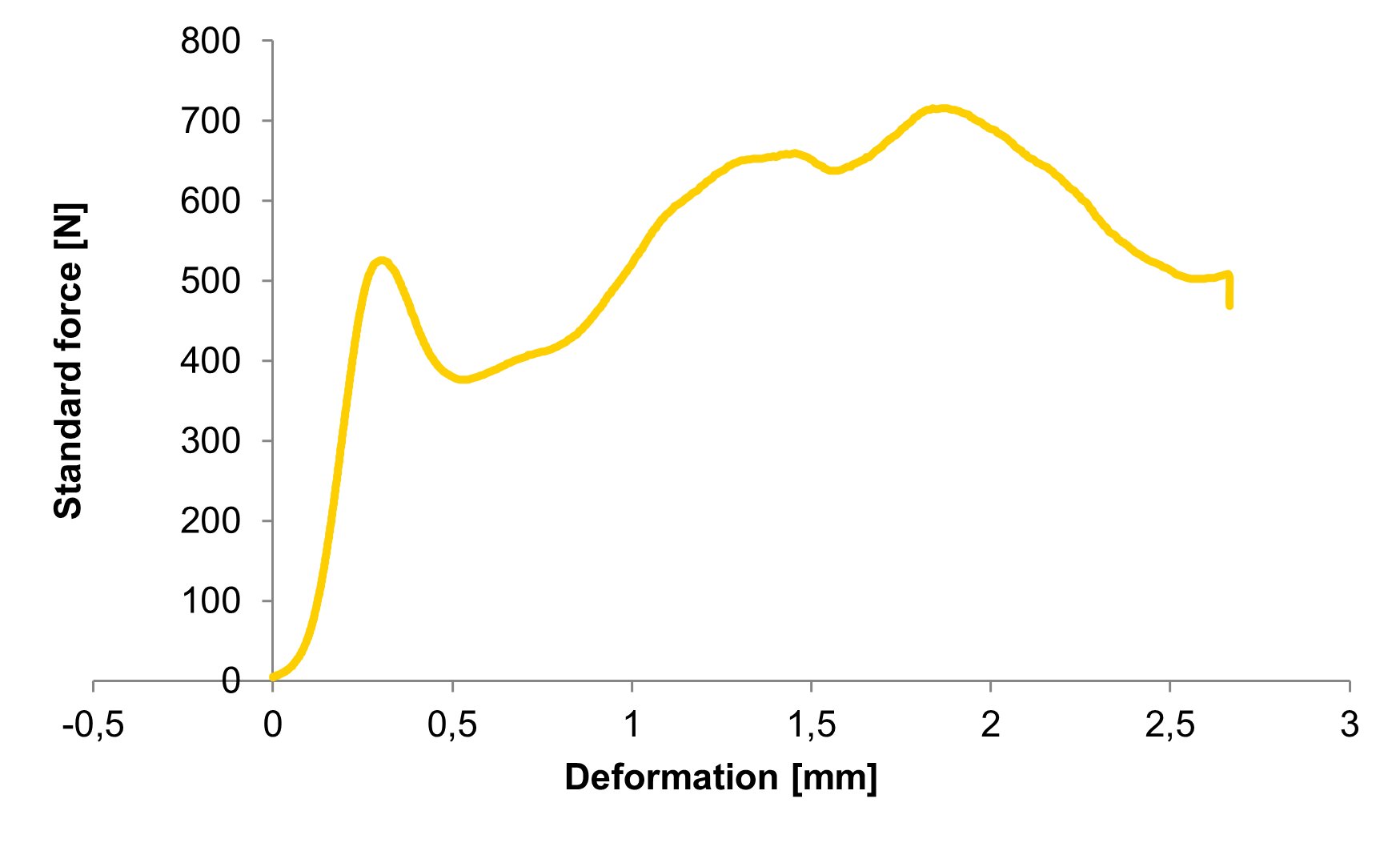}
    \caption{Load-deformation curve for the results in the analyzed sample.}
    \label{fig4}
\end{figure}

The results from the analyzed sample from the FCT are summarized in Table~\ref{tab:fct_results}. The load-deformation curve for this sample can be visualized in Figure \ref{fig4}. Determining the optimal elastic properties led to a significantly higher stiffness in the MD compared to what was determined with the ECT and SCT experiments. This difference in the results can be explained by the existence of an unknown and significant correction factor usually existing in buckling analysis. In the case of the corrugated board specimens, the thickness is not significantly lower than the remaining dimensions, meaning that the existence of such a buckling coefficient is plausible. Considering $E_{MD}$ of 1.65 MPa, $k\approx0.47$ is a valid approximation of such a correction factor, but this may vary as the paper is a stochastic material. 
\hl{It is also important to emphasize that the estimated modulus in the Z-direction of the paper material ($E_{ZD}$) is substantially lower than the in-plane moduli ($E_{MD}$ and $E_{CD}$), which is characteristic of the anisotropic behavior of paper. However, the out-of-plane effective stiffness of the full corrugated board is not determined by $E_{ZD}$ alone. Due to the sinusoidal geometry of the flutes and their orientation, the transformation of the compliance matrix results in coupling between the machine, cross, and thickness directions. As a consequence, the final out-of-plane effective stiffness ($E_{z,\text{eff}}$) reflects contributions from all three principal material directions. This coupling is intrinsic to the homogenization process applied in this study and helps to explain why the effective modulus cannot be interpreted solely based on $E_{ZD}$.}

\begin{table}[H]
    \centering
    \caption{Estimated Young's Moduli in different directions based on FCT analysis.}
    \label{tab:fct_results}
    \begin{tabular}{lc}
        \toprule
        \textbf{Variable} & \textbf{Value [MPa]} \\
        \midrule
        Experimental Effective $E$ & \textbf{5.588} \\
        Optimized $E_{MD,FCT}$ & 3509.2 \\
        Optimized $E_{CD,FCT}$ & 1330.4 \\
        Optimized $E_{Z,FCT}$ & 0.0509 \\
        \bottomrule
    \end{tabular}
\end{table}

\begin{figure}[H]
    \centering
    \includegraphics[width=0.7\textwidth]{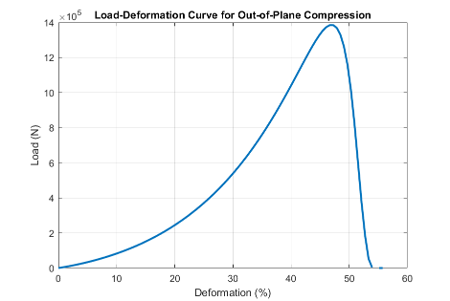}
    \caption{Load-deformation curve resulted from the homogenization process and before considering failure.}
    \label{fig5}
\end{figure}

Applying the model describing the flute deformation, using the resulting material properties and the out-of-plane homogenization methodology, the load-deformation curve considering no failure is obtained, resulting in Figure \ref{fig5}. \hl{Before achieving the peak, the fluted core undergoes deformation that causes portions of its arc to rotate such that the local machine direction (MD), which is aligned with the fiber orientation of the paper, becomes nearly vertical. When this local MD aligns with the compression load direction (global Z), the structural response temporarily stiffens, since the strongest fiber direction is aligned with the applied stress. This transient configuration occurs just before the onset of buckling, after which the flute loses stability and stiffness drops sharply.}

\begin{figure}[H]
    \centering
    \includegraphics[width=\textwidth]{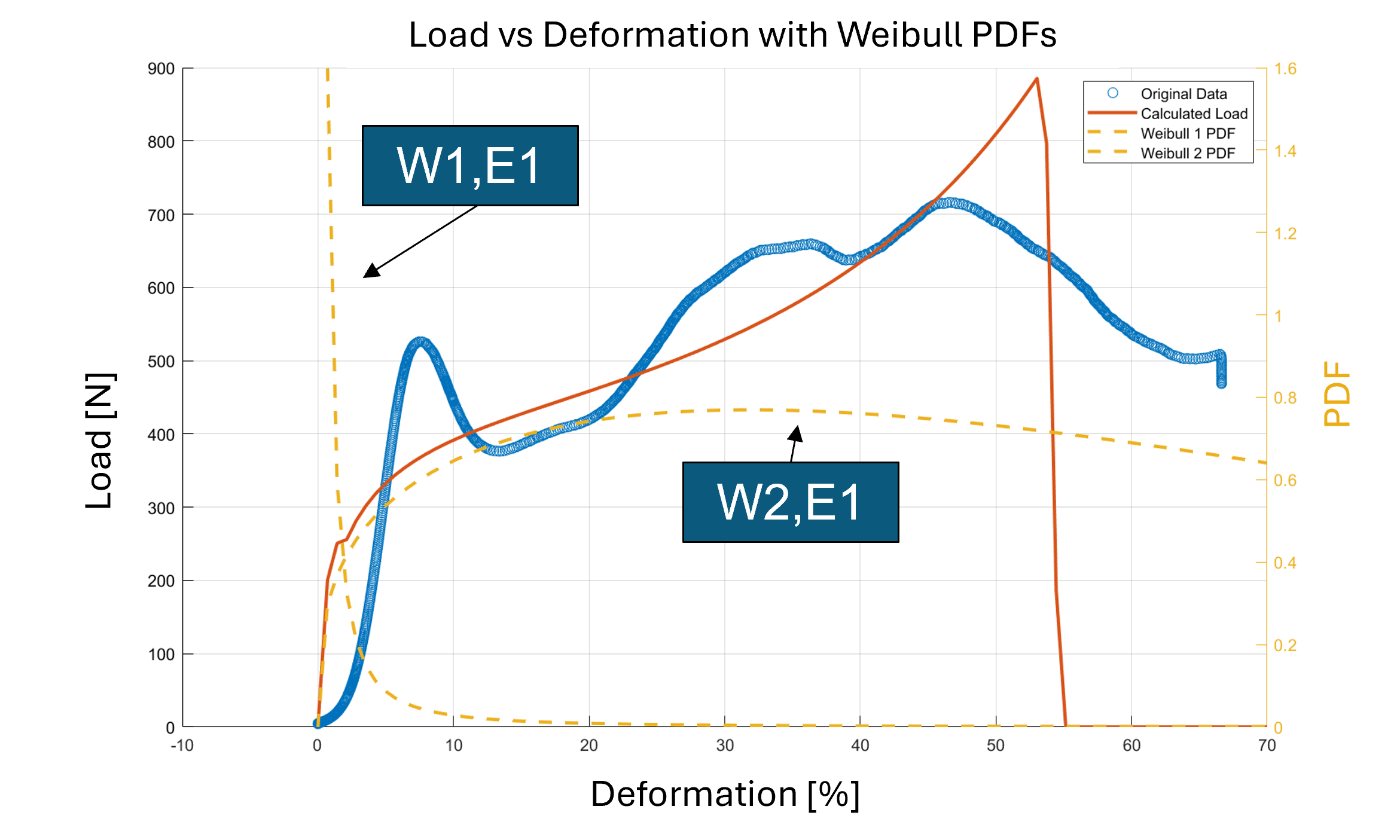}
    \caption{Load-deformation curve resulted from the application of the Weibull distributions.}
    \label{fig6}
\end{figure}

The statistical Weibull parameters derived from the experiments are shown in Table~\ref{tab:weibull_results}, and the resulting load-deformation curve is described in Figure \ref{fig6}. The shape parameter of the first curve $\beta_1=0.14$ suggests that the distribution is capturing a failure in the starting phase of the compression, which corresponds to the initial contact between liners and flute. This can explain the energy loss captured by the Weibull distribution. The shape parameter of the second curve $\beta_2=1.31$, on the other hand, suggests that the failures mostly occur randomly during the compression. This can be explained not only by the fact the buckling of all flute waves does not happen at the same time, as the flutes contain imperfections, and the distribution may be capturing them.

\begin{table}[H]
    \centering
    \caption{Weibull parameters from reference test.}
    \label{tab:weibull_results}
    \begin{tabular}{lcc}
        \toprule
        \textbf{Experiment} & \textbf{Parameter} & \textbf{Value} \\
        \midrule
        W1,E1 & $\beta_1$ & 0.14 \\
        & $\eta_1$ & 4.82e-7 \\
        W2,E1 & $\beta_2$ & 1.31 \\
        & $\eta_2$ & 0.96 \\
        \bottomrule
    \end{tabular}
\end{table}

\begin{figure}[H]
    \centering
    \includegraphics[width=\textwidth]{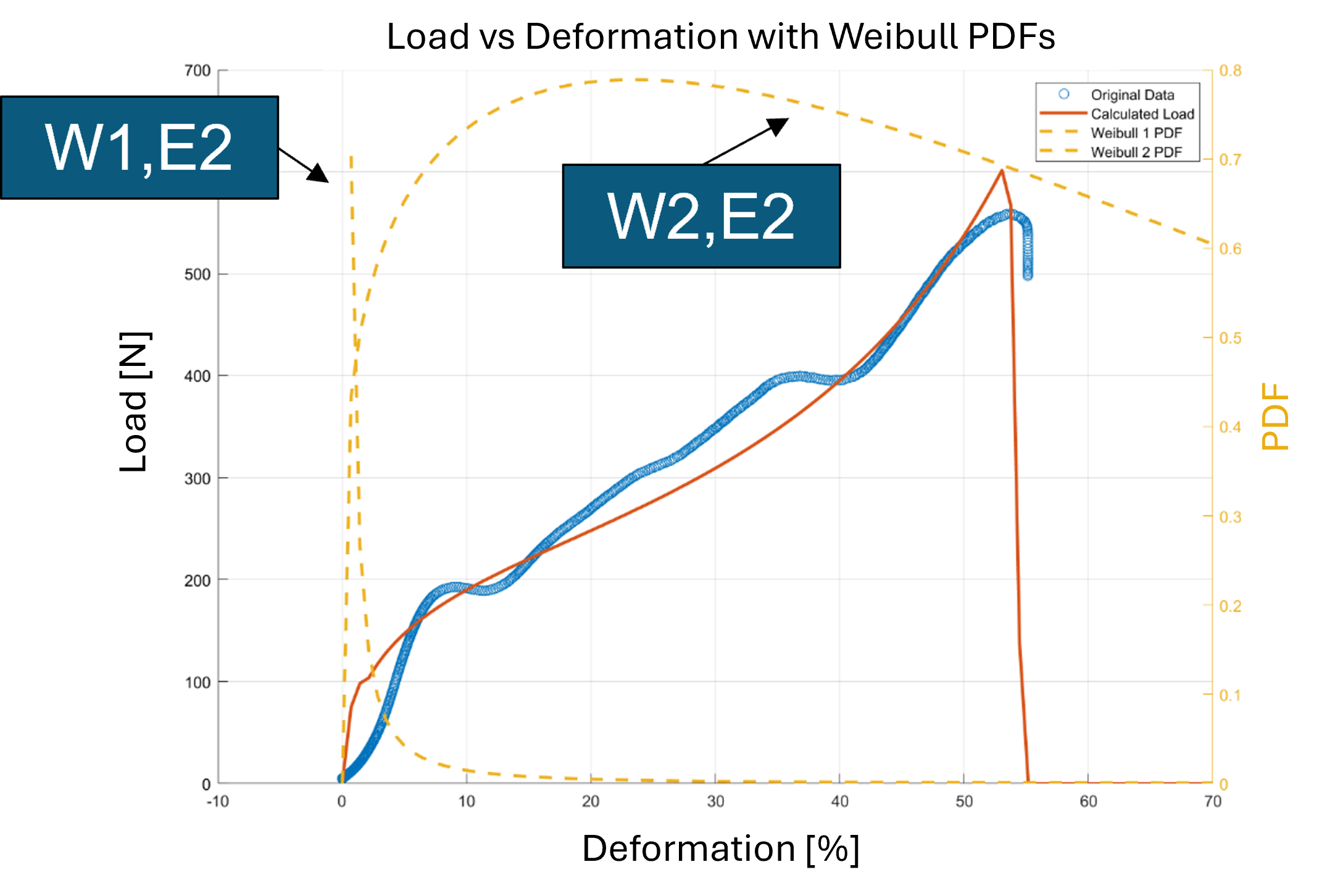}
    \caption{Load-deformation curve resulted from the application of the Weibull distributions for a second specimen.}
    \label{fig7}
\end{figure}

\hl{To validate the model, an independent specimen — drawn from the same nominal material configuration but manufactured separately — was used, and was not part of the parameter calibration process. The previously determined Weibull parameters (from Table} \ref{tab:weibull_results}) \hl{were directly compared to the new  Weibull parameters for the new experimental setup. The load-deformation curve corresponding to this new specimen compression is shown in Figure} \ref{fig7}. \hl{The corresponding Weibull parameters estimated from this second experiment (Table} \ref{tab:weibull_results2}) \hl{were also found to be in close agreement with the first, confirming the robustness of the approach. In particular,} the shape parameters of both curves ($\beta_1=0.11$ and $\beta_2=1.22$) seem to be very similar to the coefficients in the first experiment (p=1.00, LRT, Table \ref{tab:stat_tests}).

\begin{table}[H]
    \centering
    \caption{Weibull parameters from the second test.}
    \label{tab:weibull_results2}
    \begin{tabular}{lcc}
        \toprule
        \textbf{Experiment} & \textbf{Parameter} & \textbf{Value} \\
        \midrule
        W1,E2 & $\beta_1$ & 0.11 \\
        & $\eta_1$ & 4.22e-9 \\
        W2,E2 & $\beta_2$ & 1.22 \\
        & $\eta_2$ & 0.95 \\
        \bottomrule
    \end{tabular}
\end{table}

\begin{table}[H]
    \centering
    \caption{Statistical tests for distribution comparison.}
    \label{tab:stat_tests}
    \begin{tabular}{lcc}
        \toprule
        \textbf{Statistical Test} & \textbf{p-value (W1, E1 vs. E2)} & \textbf{p-value (W2, E1 vs. E2)} \\
        \midrule
        LRT* & 1.00 & 1.00 \\
        KS** & <0.05 & 0.76 \\
        Bootstrap & 0.74 & 0.53 \\
        \bottomrule
    \end{tabular}
    \small *Likelihood Ratio Test (LRT)\\
    **Kolmogorov-Smirnov (KS) Test
\end{table}

\hl{While the global deformation of the corrugated board under out-of-plane compression reaches approximately 70\%, this does not imply that the material itself experiences equally large strains. The high displacement values result primarily from structural mechanisms — including flute rotation, progressive flattening, and eventual buckling — rather than from excessive local material deformation. Consequently, the assumption of linear elasticity remains valid for describing the effective structural response up to the peak load. Post-peak nonlinearities, including energy dissipation and structural collapse, are captured statistically through the Weibull distributions. This modeling strategy enables efficient prediction while maintaining realistic representation of both elastic and failure behavior.}

\begin{figure}[H]
    \centering
    \includegraphics[width=0.5\textwidth]{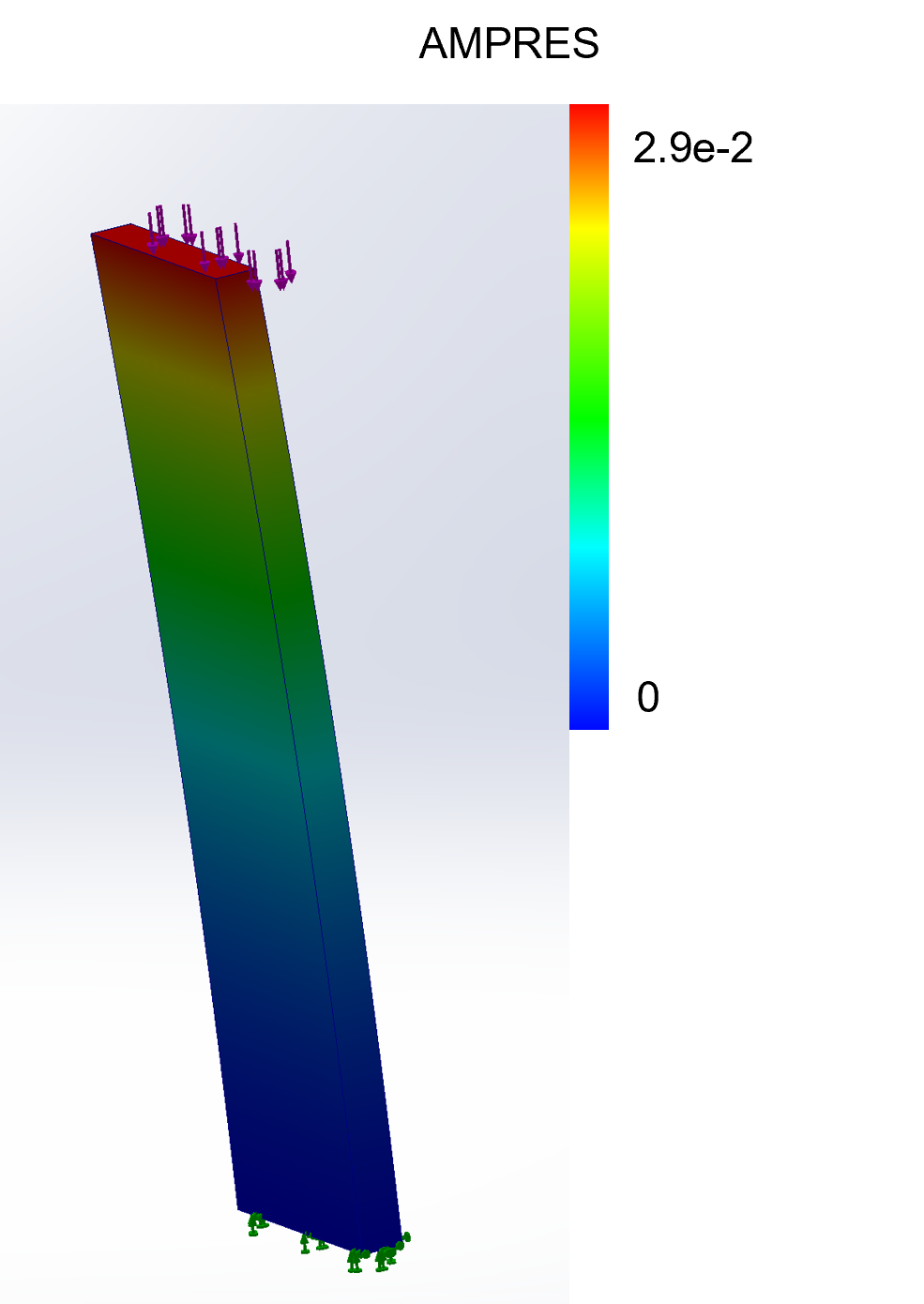}
    \caption{FEM buckling analysis considering the corrugated board as a material.}
    \label{fig8}
\end{figure}

The final FEM approach is shown in Figure~\ref{fig8}, where the statistical parameter values are combined with the effective thickness and elastic modulus obtained from the homogenization framework. A corrugated board is now represented as a single solid block, with the homogenized out-of-plane stiffness used as an input for a custom material model. \hl{In this application case, the material behavior is modeled as linear elastic, and no internal state variables or nonlinear constitutive updates are included. Consequently, the FEM simulation does not require iterative schemes such as Newton-Raphson, as the response remains within the linear regime.}

\hl{It is important to note, however, that if the material model were to be extended to include deformation-dependent stiffness, rate effects, or nonlinear degradation, an iterative solution strategy such as Newton-Raphson would indeed be necessary, even under small strain assumptions.}

\hl{The present FEM example is intended purely as a proof of concept, illustrating how the homogenized and statistically corrected material properties can be used in a structural simulation under compressive loading. While not part of the core model validation, this application demonstrates the integration potential of the proposed methodology within packaging design workflows.}

Still, the computational simplification is expected to be significant due to the reduced-order material representation introduced in this study.

\hl{Despite its effectiveness in modeling axial compression behavior, the proposed framework has several limitations. One limitation is that it assumes linear elastic behavior, with nonlinear effects and material failure introduced only via statistically calibrated Weibull distributions. As such, it does not represent damage evolution or plasticity in a mechanistic manner. The model also assumes a regular flute geometry and does not capture geometric imperfections or manufacturing variability. Furthermore, the validation has been performed for axial compression scenarios only. Further testing under multiaxial or dynamic conditions would be needed to fully assess the model’s generalizability.}

\hl{It is also acknowledged that the present model does not explicitly resolve out-of-plane shear stiffness, which is known to influence the failure behavior of corrugated board structures—particularly under bending or complex edgewise loading, as established in foundational studies such as} \parencite{carlsson2001elastic}. \hl{As such, the model is primarily applicable to configurations dominated by axial compression (e.g., FCT, simplified ECT). However, the use of statistically calibrated Weibull distributions provides a mechanism to capture the effects of failure and post-peak nonlinear behavior, including those indirectly related to shear-induced instabilities. These statistical corrections, applied to the homogenized elastic baseline, allow the model to approximate complex failure mechanisms without requiring explicit geometric or material nonlinearity. Furthermore, the homogenization process itself builds upon and extends existing methodologies in the literature by incorporating an orientation-aware compliance transformation, a layer-wise integration approach, and an inverse calibration strategy directly informed by experimental data. Still, the influence of shear is considered secondary and was not independently evaluated. Future work may extend the framework to explicitly calibrate and validate shear-dominated behavior for bending, torsion, or box compression scenarios.}

\subsection{Conclusions and Future Work}
The study successfully developed a homogenization-based FEM model for corrugated boards. The methodology provided estimated predictions of mechanical properties compared to experimental data, demonstrating the effectiveness of the homogenization process and the importance of considering material heterogeneity.

The incorporation of Weibull distributions proved essential for capturing both early and progressive failure modes, allowing for a more comprehensive understanding of material behavior. This statistical approach ensures that the failure mechanisms are better accounted for in future structural analyses.

Future research will focus on expanding the model to better analyze the statistical parameters of the Weibull distribution. Additionally, alternative flute geometries and compositions will be explored to evaluate their impact on mechanical performance and failure behavior. \hl{Vertical loading, particularly under bending, induces significant interfacial shear stresses that may contribute to delamination. While such mechanisms are not explicitly resolved in this model, they are implicitly considered within the out-of-plane Weibull distribution. A more detailed shear-driven delamination model could be explored in future work.} Finally, the integration of machine learning models for faster predictions and optimization will be investigated to further enhance design efficiency and to make the statistical parameters more AI-driven for customization.

\section*{Ethical Declaration}

The authors acknowledge systemic accountability issues, namely that access to troubleshooting data from the testing machine was limited, restricting independent verification of potential machine malfunctions. However, the machine expert has confirmed that the reported data was not affected, ensuring the reliability and validity of the experimental results. This limitation makes it necessary for systemic accountability measures to enhance transparency and reproducibility.

\section*{Note}
This manuscript has been submitted for consideration for publication in the TAPPI Journal. The content is under review and has not yet been fully peer-reviewed or accepted.

\printbibliography

@mastersthesis{schmitt2019fibre,
  author    = {Schmitt, C.},
  title     = {Investigation of the influence of fibre sources on the strength of corrugated boxes and prediction of final strength properties from basic furnish properties},
  school    = {Technische Universität Darmstadt},
  year      = {2019},
  address   = {Darmstadt, Germany}
}

@article{nordstrand1994transverse,
  title={Transverse shear stiffness of structural core sandwich},
  author={Nordstrand, Tomas and Carlsson, Leif A and Allen, Howard G},
  journal={Composite structures},
  volume={27},
  number={3},
  pages={317--329},
  year={1994},
  publisher={Elsevier}
}

@article{carlsson2001elastic,
  title={On the elastic stiffnesses of corrugated core sandwich},
  author={Carlsson, Leif A and Nordstrand, Tomas and Westerlind, BO},
  journal={Journal of Sandwich Structures \& Materials},
  volume={3},
  number={4},
  pages={253--267},
  year={2001},
  publisher={Sage Publications Sage CA: Thousand Oaks, CA}
}

@book{nyman2004continuum,
  author    = {Nyman, U.},
  title     = {Continuum mechanics modelling of corrugated board},
  year      = {2004}
}

@misc{DIN55468-1_2021,
  title     = {Packstoffe Wellpappe - Teil 1: Anforderungen, Prüfung (DIN 55468-1:2021-03)},
  author    = {{DIN Deutsches Institut für Normung e.V.}},
  year      = {2021},
  note      = {Berlin: Beuth Verlag},
}

@misc{iso9895,
  author    = {{International Organization for Standardization}},
  title     = {ISO 9895: Paper and board — Compressive strength — Short-span test},
  year      = {2004}
}

@misc{iso3037,
  author    = {{International Organization for Standardization}},
  title     = {ISO 3037: Corrugated fiberboard — Determination of edgewise crush resistance (ECT)},
  year      = {2004}
}

@phdthesis{starke2020corrugated,
  author    = {Starke, M. M.},
  title     = {Material and structural modelling of corrugated paperboard packaging for horticultural produce},
  school    = {Stellenbosch University},
  year      = {2020},
  address   = {Stellenbosch, South Africa}
}

@misc{iso3035,
  author    = {{International Organization for Standardization}},
  title     = {ISO 3035: Corrugated fiberboard — Determination of flat crush resistance},
  year      = {2004}
}

@article{marek2017homogenization,
  author    = {Marek, A. and Garbowski, T.},
  title     = {Homogenization of sandwich panels},
  journal   = {Computer Assisted Methods In Engineering And Science},
  volume    = {22},
  number    = {1},
  pages     = {39-50},
  year      = {2017},
  url       = {https://cames.ippt.gov.pl/index.php/cames/article/view/13}
}

@article{fitas2023review,
  title={A Review of Optimization for Corrugated Boards},
  author={Fitas, Ricardo and Schaffrath, Heinz Joachim and Schabel, Samuel},
  journal={Sustainability},
  volume={15},
  number={21},
  pages={15588},
  year={2023},
  publisher={MDPI}
}

@inproceedings{fitas2025corrugated,
  author    = {Fitas, R.},
  title     = {Modeling the effective elastic modulus and thickness of corrugated boards using Gaussian process regression and expected hypervolume improvement},
  booktitle = {Proceedings of International Conferences on Digital Technology Driven Engineering},
  publisher = {Springer},
  year      = {2025},
  note      = {In press}
}

@article{kerfriden2014certification,
  title={Certification of projection-based reduced order modelling in computational homogenisation by the constitutive relation error},
  author={Kerfriden, Pierre and R{\'o}denas, Juan-Jos{\'e} and Bordas, SP-A},
  journal={International Journal for Numerical Methods in Engineering},
  volume={97},
  number={6},
  pages={395--422},
  year={2014},
  publisher={Wiley Online Library}
}

@article{russo2023evaluation,
  title={Evaluation of wave configurations in corrugated boards by experimental analysis (EA) and finite element modeling (FEM): the role of the micro-wave in packaging design},
  author={Russo, F. M. D. and Gisario, A. and Barletta, M.},
  journal={The International Journal of Advanced Manufacturing Technology},
  volume={124},
  pages={123--134},
  year={2023},
  doi={10.1007/s00170-023-11397-y}
}

@article{grabski2023identification,
  title={Identification of Geometric Features of the Corrugated Board Using Images and Genetic Algorithm},
  author={Grabski, J. K. and Garbowski, T.},
  journal={Sensors},
  volume={23},
  number={13},
  pages={6242},
  year={2023},
  doi={10.3390/s23136242}
}

@article{cornaggia2023influence,
  title={Influence of humidity and temperature on mechanical properties of corrugated board - Numerical investigation},
  author={Cornaggia, A. and Gajewski, T. and Knitter-Piątkowska, A. and Garbowski, T.},
  journal={BioResources},
  volume={18},
  number={4},
  pages={7490--7509},
  year={2023},
  doi={10.15376/biores.18.4.7490-7509}
}

@article{aduke2023analysis,
  title={An Analysis of Numerical Homogenisation Methods Applied on Corrugated Paperboard},
  author={Aduke, R. N. and Venter, M. P. and Coetzee, C. J.},
  journal={Mathematical and Computational Applications},
  volume={28},
  number={2},
  pages={46},
  year={2023},
  doi={10.3390/mca28020046}
}

@article{park2024finite,
  title={Finite Element-based Simulation for Bending Behavior of Corrugated Package},
  author={Park, J. M. and Kim, J. S. and Sim, J. M. and Jung, H.-M.},
  journal={Preprints},
  year={2024},
  doi={10.20944/preprints202408.1902.v1}
}

@article{cillie2022experimental,
  title={Experimental and Numerical Investigation of the In-Plane Compression of Corrugated Paperboard Panels},
  author={Cillie, J. and Coetzee, C.},
  journal={Mathematical and Computational Applications},
  volume={27},
  number={6},
  pages={108},
  year={2022},
  doi={10.3390/mca27060108}
}

@article{greco2009homogenized,
  title={Homogenized mechanical behavior of composite micro-structures including micro-cracking and contact evolution},
  author={Greco, F.},
  journal={Engineering Fracture Mechanics},
  volume={76},
  number={8},
  pages={1211--1233},
  year={2009},
  doi={10.1016/j.engfracmech.2008.09.006}
}

@incollection{wriggers2007homogenization,
  title={Homogenization and Multi-Scale Approaches for Contact Problems},
  author={Wriggers, P. and Nettingsmeier, J.},
  booktitle={Computational Contact Mechanics},
  pages={129--161},
  year={2007},
  publisher={Springer},
  doi={10.1007/978-3-211-77298-0_4}
}

\end{document}